# Feasibility of Haralick's Texture Features for the Classification of Chromogenic In-situ Hybridization Images

Stoyan Pavlov
*Dept. of Anatomy and cell biology and Advanced Computational Bioimaging, Research Institute*
*Medical University Varna*
*Medical University "Prof. Dr Paraskev Stoyanov"*
Varna, Bulgaria
ORCID: 0000-0002-9322-2299

Galina Momcheva
*Dept. of Computer Science*
*Varna Free University "Chernorizets Hrabar"*
*Advanced Computational Bioimaging, Research Institute*
*Medical University "Prof. Dr Paraskev Stoyanov"*
Varna, Bulgaria
galina.momcheva@vfu.bg

Pavlina Burlakova
*Dept. of Computer Science*
*Varna Free University "Chernorizets Hrabar"*
Varna, Bulgaria
172831003@vfu.bg

Simeon Atanasov
*Dept. of Computer Science*
*Varna Free University "Chernorizets Hrabar"*
*High School of Mathematics "Dr. Petar Beron"*
Varna, Bulgaria
simeon.a.atanasov@gmail.com

Dimo Stoyanov
*Dept. of Anatomy and cell biology and Advanced Computational Bioimaging, Research Institute*
*Medical University "Prof. Dr Paraskev Stoyanov"*
Varna, Bulgaria
ORCID:0000-0002-2324-0002

Martin Ivanov
*Dept. of Anatomy and cell biology and Advanced Computational Bioimaging, Research Institute*
*Medical University "Prof. Dr Paraskev Stoyanov"*
Varna, Bulgaria
martin.ivanov@mu-varna.bg

Anton Tonchev
*Dept. of Anatomy and cell biology*
*Medical University "Prof. Dr Paraskev Stoyanov"*
Varna, Bulgaria
anton.tonchev@mu-varna.bg

*Abstract*—This paper presents a proof of concept for the usefulness of second-order texture features for the qualitative analysis and classification of chromogenic in-situ hybridization whole slide images in high-throughput imaging experiments. The challenge is that currently, the gold standard for gene expression grading in such images is expert assessment. The idea of the research team is to use different approaches in the analysis of these images that will be used for structural segmentation and functional analysis in gene expression. The article presents such perspective idea to select a number of textural features that are going to be used for classification. In our experiment, natural grouping of image samples (tiles) depending on their local texture properties was explored in an unsupervised classification procedure. The features are reduced to two dimensions with fuzzy c-means clustering. The overall conclusion of this experiment is that Haralick features are a viable choice for classification and analysis of chromogenic in-situ hybridization image data. The principal component analysis approach produced slightly more "understandable" from an annotator's point of view classes.

## I. Introduction

High-throughput chromogenic in-situ hybridization (CISH) is an invaluable technique to study the spatial distribution of gene expression [1]. The method is used to detect known mRNA-sequences using specific complementary hybridization with antisense probes tagged with enzyme-linked haptens. In the chromogenic variant of the reaction, the hybridization is revealed by the production of colored precipitate that can be easily observed, recognized, and documented using a standard bright-field microscopic imaging system. CISH is primarily used to localize the specific mRNA fragments of the sought gene product and the cell that produces it in fixed tissues [2], [3]. The positive signals correspond to cells that actively produce (aka express) the product of a gene under investigation.

This paper presents a proof of concept for the usefulness of second-order texture features for the qualitative analysis and classification of CISH whole slide images in high-throughput imaging experiments. The paper is organized as follows: In section 2. is explained the rationale and motifs behind the proposed approach. Section 3 discusses the selected texture features. In section 4 is outlined the experiment and its results. In section 5 we discuss the evaluation of the domain-experts. In section 5 we conclude with some remarks on further developments.

## II. Rationale

Currently, the gold standard for gene expression grading in the CISH stained tissue slides is expert assessment. This approach involves visual inspection of the cell expression and manual labeling (positive or negative) or grading depending on the amount of the observed precipitate. The usual visual scoring is based on cellular gene expression strength ("negative", "low", "moderate", or "strong"), and patterns of expression ("ubiquitous", "regional" or "scattered") [4], [5]. This method is highly biased and suffers from low reproducibility as it strongly depends on the conditions and the expertise of the annotator. Furthermore, it is slow and is not particularly effective in high-throughput experiments, when large amounts of image data are generated at high speeds. There are several attempts to design automated workflows for an unbiased evaluation of gene expression. Celldetect is an open-source algorithm for automatic localization and grading of cellular gene expression whole-slide CISH images. [6]. The algorithm performs intensity-based thresholding, classification, and labeling at a reduced spatial resolution (approximately single cell per pixel). While this approach significantly reduces bias and improves reproducibility, it still suffers from the high variability between images and staining batches inherent to the method. The good reproducibility of Celldetect relies heavily on the human operator to select the proper parameters to account for intensity variability between images and batches. Another reliable quantitative workflow is developed by Allen Brain Institute and is implemented in the annotation of their Mouse Brain ISH Atlas [7], [8]. The workflow evaluates gene expression in high-resolution images with normalized brightness based on the grey value of the corresponding pixels in combination with advanced filtering to account for different patterns of expression [9], [10]. Both approaches are very effective but are sensitive to brightness fluctuations and noise introduced by the ISH procedure or the image acquisition, and require very strict control of experimental conditions.

A robust and automated workflow can facilitate reproducibility between experiments and between labs and can ensure the acquisition of comparable data from imaging CISH experiments.

During manual evaluation experienced annotators are able to recognize similar levels of gene expression despite of significant differences in the overall brightness between images. This is achieved by an implicit evaluation of brightness and color distribution within the locality of the observed tissue. The most obvious property of an image that is related to the local fluctuations of intensity and color is the texture. Recent research showed that second-order textural features can be used successfully to classify and localize gene expression patterns to specific cerebellar cortical layers [11] or to identify mRNA enriched sites in the hippocampal region of the brain [12].

### III. TEXTURE FEATURES

In the image processing, the texture is a repeating pattern or a function of spatial variation of the brightness intensity of the pixels. Texture analysis plays an important role in face recognition, surface defect detection, pattern recognition, and medical image analysis by the extraction of meaningful information from digital images. Texture analysis includes statistical, structural, model-based and transform methods. The textures in images are not uniform in color, scale, and orientation that is why we are searching for specific characteristics in images. Texture feature extraction refers to the process of computing characteristics of image which numerically describes textural image properties and may include scalar numbers, discrete histograms or empirical distributions [13].

Generally, CISH images have inhomogeneous texture. That is why a particular standard method for feature extraction is not enough for their analysis and a combination of features are exposed. The idea of the research team is to use different approaches in the analysis of CISH images that will be used for structural segmentation and functional analysis in gene expression. The article presents such perspective idea to select a number of textural features that are going to be used for classification.

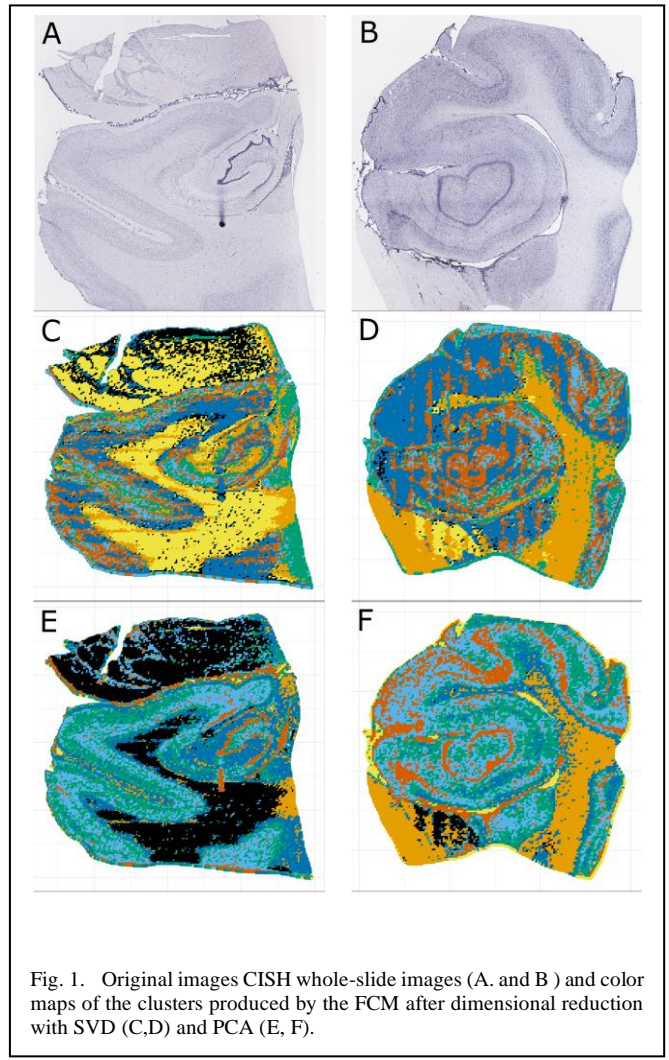

Fig. 1. Original images CISH whole-slide images (A. and B) and color maps of the clusters produced by the FCM after dimensional reduction with SVD (C,D) and PCA (E, F).

The Haralick features are extracted from the Grey-Level Co-occurence Matrix (GLCM) that is a well-known statistical technique for selection of second order statistics of an image. This group of statistics accounts for the spatial inter-dependency of two pixels at specific relative positions. Haralick et al. [14] suggested a set of 14 textural features which can be extracted from the co-occurrence matrix, and which contain information about image textural characteristics such as homogeneity, linearity, and contrast. They include the variance calculated on the sum of adjacent pixels; variance on the difference between adjacent pixels; entropy on the sum and on the difference; correlation involving entropies, and the maximum correlation coefficient. After evaluating possible texture analysis features methods and approaches as Gabor [15] and Haralick features [16] and after some preliminary research work with GLCM and the Gray Level Histogram (GLH) the team decided to first focus on experiments with Haralick features.

### IV. THE EXPERIMENT

The natural grouping of image samples (tiles) depending on their local texture properties was explored in an unsupervised classification procedure. The experiment was performed in the following steps:

Test images. CISH stained whole-slide scanned images (spatial resolution - 0.5 μm/px) from the hippocampal region of a primate brain provided by the Department of anatomy and cell biology at the Medical University - Varna. The riboprobe visualization system uses BCIP/NBT-substrate as a chromogen, that precipitates and stains the riboprobes in dark blue to purple color. (Fig.1A,B)

Measurement. The images were opened in QuPath v.0.2.1 (Bankhead et al., 2017). Using the annotation tools of the software the tissue slices were selected from the surrounding background and all subsequent measurements were performed only on the selections. The selected tissue was scanned with overlapping circular regions (size = 150 μm, step = 100 μm, overlap = 25 μm). In each circular neighborhood, the Haralick features based on the grey level co-occurrence matrix (distance = 1px and bins = 127 ) were measured for each overlapping tile. A total of 13 Haralick features were calculated in the "Brightness" and "Saturation" channels of the HSB color space, forming 26-dimensional feature vector - Angular second moment (F0), Contrast (F1), Correlation (F2), Sum of squares (F3), Inverse difference moment (F4), Sum average (F5), Sum variance (F6), Sum entropy (F7), Entropy (F8), Difference variance (F9), Difference entropy (F10), Information measure of correlation 1 (F11) and Information measure of correlation 2 (F12).

Dimensionality reduction. Two techniques were tested separately - Singular Value Decomposition - SVD [17], [18] and Principal Component Analysis - PCA [18], [19]. This and all subsequent steps were programmed in Python Programming Language (Python Software Foundation, https://www.python.org/) and libraries in the SCIPy ecosystem [20]–[23]. The purpose of this step was to eliminate noisy data dimensions and thus improve accuracy in clustering, in addition to the reduced computational cost [24], [25]. In both cases, the features were reduced to two dimensions. On one hand, a high number of dimensions seem to influence the fuzzy c-means clustering (FCM) algorithm negatively. From a different point of view, the annotator is evaluating the CISH images mainly on two properties - strength of expression and pattern of expression (density). Thus it seemed appropriate to keep the first two components for further analysis.

Fuzzy C-means clustering. A Fuzzy C-Means (FCM) was implemented for the quantitative clustering of the tiles in the described images. Our preference for the fuzzy counterpart of the more popular k-means algorithm was driven mainly by the notion that in living organisms groups and categories are fuzzy by nature and an approach that captures this fuzziness will be more beneficial. For the selection of a cluster number, we looped the FCM between two and ten cluster centers on the two components generated via SVD and PCA separately and evaluated the results with the fuzzy partition coefficient [26]. While the FPC reaches a maximum value at two clusters, our choice was once again based on the nature of the expert evaluation of the target images: the standard protocol for evaluation includes at least four (and if pattern is taken into account up to eight separate groupings). Thus, we chose seven clusters - a value that would allow us to capture gradual differences in gene expression and at which the FPC is still at acceptable values. Upon parameterization, it was ensured that the results are set in a way where final clusters appear in the same order. We calculated the centroid for each cluster and for each point, compute its coefficients of being in the clusters [27]

V. EXPERT EVALUATION.

To characterize the generated classes we sent the original unprocessed images and the classification maps to four evaluators to assess the strength of gene expression in each class ("none", "low", "moderate" or "strong"; score from 0 to 3) and its pattern ("none", "sparse" or "dense"; score from 0 to 2). In a next run the annotators received ten random tiles from each class (altogether 70 tiles per image) and classified them according to the accepted scheme. Each evaluator received a weight depending on their expertise – Evaluator 1 – weight 3, Evaluator 2 – weight 2, Evaluators 3 and 4 – weight 1. The final scores for each class and tile were received as a weighted mean of the four separate evaluations. The final class labels were compared to the tile classifications by the domain experts and confusion matrices were generated (Table 1).

The fuzzy clustering of the reduced data produced classes that on inspection separated the images into regions with different expression levels (Fig. 1). Even for the unprepared observer the contrast distribution in the unprocessed images corresponds to the cluster mapping.

The overall accuracy is relatively low, but there is a lot of noise due to misclassifications in close evaluation levels (i.e. "medium" and "high"). When looking at the confusion matrices of the separate evaluation levels we can see that both approaches give a relatively good accuracies. For such low level and biased aproach.

VI. CONCLUSION.

The overall conclusion of this experiment is that Haralick features are a viable choice for classification and analysis of CISH image data. The PCA approach produced slightly more "understandable" from an annotator's point of view classes.

Of course there are a lot of additional problems to solve. First of all, the homogeneity, contrast, entropy and energy are sensitive to the choice of the direction. The selected images for this research while not uniform and isotropic are not influenced by this restriction conditions due to the fact that at the chosen scale of chosen tiles the images become isotropic. Many of the features correlate with each other and probably that is why PCA worked a little bit better as it reduces the importance of non-correlating features on the principle components. Next step should be to perform an exploratory factor analysis of these features and select only the most important ones.

An extension to this set with additional calculable features that measure frequency and orientations is a good approach. Another good candidate for this are Gabor filters, that represent frequencies and orientation in a way very similar to human vision, and thus may give us a tool to quantify the currently implicit and unquantifiable decision of the domain expert, that is - e.g. a measurable repeatable statistic that corresponds to "highly expressed" gene with "medium density" .

ACKNOWLEDGMENT

The research is conducted under the supervision of Assoc. Prof. Stoyan Pavlov and Assoc. Prof. Galina Momcheva leading research groups under the R&D ecosystem Biomed

Varna (www.biomedvarna.com). Prof. Pavlov and Prof. Momcheva contributed equally and share senior authorship.

This work is partially supported by the projects "University - Varna Science Fund project 19028" and Bulgarian Academy of Sciences funding 2019 for jVarnaBioImage project with participating students from High School of Mathematics – Varna.